\documentclass[runningheads]{llncs}
\usepackage[T1]{fontenc}
\usepackage{graphicx}
\begin{document}
\title{Distribution Grid Monitoring Based on Widely Available Smart Plugs}
\author{Simon Grafenhorst\inst{1}\orcidID{0000-0002-6366-0626} \and
Kevin Förderer\inst{1}\orcidID{0000-0002-9115-670X} \and
Veit Hagenmeyer\inst{1}\orcidID{0000-0002-3572-9083}}
\authorrunning{S. Grafenhorst et al.}
\institute{Institute for Automation and Applied Informatics (IAI), Karlsruhe Institute of Technology, Hermann-von-Helmholtz-Platz 1, 76344, Eggenstein-Leopoldshafen, Germany}
\maketitle              % typeset the header of the contribution
\begin{abstract}
The growing popularity of e-mobility, heat pumps, and renewable generation such as photovoltaics is leading to scenarios which the distribution grid was not originally designed for. Moreover, parts of the distribution grid are only sparsely instrumented, leaving the distribution system operator unaware of possible bottlenecks resulting from the introduction of such loads and renewable generation.
To overcome this lack of information, we propose the use of widely available smart home devices, such as smart plugs, for grid monitoring. We detail the aggregation and storage of smart plug measurements for distribution grid monitoring and examine the accuracy of the measurements. A case study shows how the average monitoring error in a distribution grid area decreases the more measurement devices are installed. Hence, simple smart plugs can help with distribution grid monitoring and provide valuable information to the DSO.

\keywords{Smart Grid \and Distribution Grid Monitoring \and Smart Home Measurement Device}
\end{abstract}
\section{Introduction}
\subsection{Motivation}
With new loads such as electric vehicle (EV) chargers and new generators such as small PV systems, the demands on the grid are changing rapidly. For example, the distribution grid can only accommodate a certain amount of photovoltaic generation without violating regulatory constraints (hosting capacity\cite{fatima_review_2020,mulenga_review_2020}). Other challenges include the optimized control of distributed energy resources (DER), such as controllable heat pumps or schedulable electric vehicle charging infrastructure, as well as demand-side management (DSM) of smart household appliances. As a result, it is essential that the distribution system operator (DSO) has comprehensive and up-to-date measurement data from the distribution grid. However, the adoption of smart distribution grid infrastructure to monitor the live state of the grid or detect bottlenecks is slow~\cite{bundesnetzagentur_fur_elektrizitat_gas_bericht_2022}. Many old network infrastructures are still in use. It is neither monitored nor equipped with the necessary communication infrastructure, and upgrading the infrastructure is costly and time-consuming. Key components in the distribution grid such as transformers and cables, have a life expectancy of about 35 years~\cite{behi_distribution_2017}, and monitoring the state of the grid with such old equipment is therefore a challenge. However, smart home devices are becoming increasingly popular. In the US, 35.8~\% and in the EU, 23.0~\% of households had smart home systems installed in 2021~\cite{martin_backman_smart_2022}. Smart plugs can turn power outlets on and off remotely, and some smart plugs also include hardware to measure the power consumption of the connected device and the line voltage of the socket outlet they are plugged into. We hypothesize that sufficiently accurate grid monitoring is possible with the use of smart plugs as measurement devices. This raises two further research questions: What is the accuracy of the measurements and can the accuracy be further improved with software modifications tailored to the voltage measurements?\\
%In this paper, we study the measurement accuracy of smart plugs and propose a system architecture to securely collect measurements from smart plugs and store them in a database. We show how a firmware modification improves the accuracy and the frequency of the measurements. A case study illustrates the correlation between the number of metering devices in the grid and the error in distribution grid monitoring, and shows that even inaccurate smart home measurement devices allow for accurate monitoring.

The contributions of this paper are as follows:
\begin{enumerate}
    \item It is shown that the measurement inaccuracies of the widely available smart plugs are low enough to be comparable to other distribution grid measurement devices.
    \item It is demonstrated how a modified firmware can increase the measurement accuracy and frequency, and the firmware version is released as open source~\cite{simon_grafenhorst_fast_2023}.
    \item A case study is presented to outline the practicality of using distributed and non-calibrated measurement devices for distribution grid monitoring.
\end{enumerate}

%We show that the measurement inaccuracies of the widely available smart plugs are low enough to be comparable to other distribution grid measurement devices. Furthermore, we show how a modified firmware can increase the measurement accuracy and frequency, and we release this new firmware version as open source. The communication and storage back-end used in this paper is scalable through the use of containerized applications and can accommodate many measurement devices. We use end-to-end encryption between all services and devices and adhere to modern security standards. Since the hardware of the commercially available smart plugs is not modified, it is safe to let residents within a distribution grid area install the plugs themselves. The grid monitoring solution proposed in this paper could be deployed by a distribution grid operator in the real world and is capable of providing a live overview of the current state of the grid. A case study outlines the practicality of using distributed and non-calibrated measurement devices for distribution grid monitoring.\\
%Compared to previous work on distributed monitoring infrastructure, the cost and ease of installation of the commercially available smart plugs is unparalleled. The expandable, containerized system design allows for the integration of new equipment from other vendors and is another unique feature.

The paper is organized as follows: Section~2 summarizes related works on distribution grid monitoring, the development of measurement devices, and the secure communication with Internet of Things (IoT) devices. The method for selecting a type of smart plug to measure voltages and analyzing the data is presented in Section~3. The measurement error of the smart plugs with an unmodified open source firmware as well as with a firmware version tailored for voltage measurements is evaluated in Section~4. Section~5 presents a case study that illustrates a realistic use case for distribution grid monitoring using smart plugs. The case study is conducted using a grid simulation and a simulation of the smart plugs with the same measurement error as the real devices shown in the evaluation. The results of the case study and the practical applicability of this research are discussed in Section~6, followed by a final conclusion in Section~7.
\section{Related Work}
\subsection{Distribution Grid Monitoring}
To assess the state of the distribution grid, several articles identify accurate voltage measurements at different nodes in the distribution grid as an important prerequisite~\cite{fatima_review_2020,andreas_abart_power_2011}. The p.u. (per unit) value describes the factor between the real voltage and the nominal voltage. There are different standards that define the minimum and maximum p.u. values for different countries. For example, the EN-50160 standard specifies a p.u. of 0.9 to 1.1 as the permissible voltage variation. Therefore, to evaluate the hosting capacity for PV systems in a part of the distribution grid, the minimum and maximum p.u. levels that occur within a predefined period of time need to be known, and thus voltage measurements are needed.\\

\subsection{State of the Art}
In the past, the lack of measurement hardware in the distribution grid led to the exploration of simulations based on sparse measurement data and pseudo measurements~\cite{dahale_sparsity_2020,primadianto_review_2017,akrami_sparse_2022}. Others have increased the observability of the distribution grid by integrating smart meter data into a state estimation~\cite{wang_review_2019,khan_smart_2022,alimardani_using_2014,jia_state_2013}. This enables the generation of a forecasts~\cite{quilumba_using_2015,samarakoon_use_2011} or the detection and localization of faults in the grid~\cite{trindade_low_2017,araujo_decision_2019}. Furthermore, network topology reduction techniques can be applied to carry out a grid state estimation with a limited number of smart meters~\cite{khan_smart_2022}. To increase the accuracy of the state estimation in~\cite{alimardani_using_2014}, the unsynchronized measurements of multiple smart meters are filtered and only the most recent measurements are included. Compared to a state estimation that assumes all smart meter measurements are recorded at the same time, the proposed method is more accurate~\cite{alimardani_using_2014}. However, the smart meters require a professional electrician to install, and the majority of meters only take measurements every 15~minutes~\cite{wang_review_2019}. In comparison, the commercially available smart plugs can be installed by anyone and measure the voltage every second with a modified firmware.\\
Leveraging the Advanced Metering Infrastructure (AMI) already present in the distribution grid saves costs and expenditures at the expense of the timeliness of the data~\cite{sanchez_observability_2017}, and consequently the accuracy of the grid state estimation at the present time.
The lack of measurement hardware also leads to inaccurate load modeling of the distribution grid transformer. To calculate load profiles of the transformer and determine whether new loads could overload the current hardware, AMI can be included in the analysis~\cite{lo_transformational_2014}. Installing monitoring devices on all transformers could also solve this problem, but is not cost effective~\cite{lo_transformational_2014}. To improve the grid model and more accurately estimate the transformer peak load, several other sources of information such as temperature, geographic, customer, and facility management data can also be included. The near real-time optimization of the distribution network with smart grid technology is identified as a significant improvement for the efficient operation of the grid~\cite{lo_transformational_2014}.\\
A smart plug to monitor voltage and frequency in real-time is designed in~\cite{dimitriou_voltage_2014}. The measured values are sent to a smartphone that is connected via Bluetooth. The smartphone then forwards the data to a web server. With their implementation, they demonstrate the feasibility of measuring the voltages at different points in the distribution grid and estimating the live state of the grid based on these measurements. The device is considered a working proof of concept for a low-cost substitute for smart metering hardware, although no measurement accuracy or time delay is specified. Other authors propose the use of specialized voltage meters to monitor the state of the grid~\cite{andreas_abart_power_2011}. They synchronize their measurements and analyze the grid state with load flow simulations based on a series of snapshots of the grid. Furthermore, the underlying grid model is extended by learning from the differences between the calculated and measured voltages at different nodes. In~\cite{ganu_nplug_2012}, a smart plug is designed for DSM. They develop a software that switches the connected load on or off depending on the voltage level and show that the load peaks are shaved off when the designed smart plugs are widely distributed in the grid. However, no communication mechanism is implemented, so the measured values cannot be used for a distribution grid monitoring. The hardware is also a prototype design that is not commercially available. To monitor meteorological variables and PV generation, a low-cost data logger device with LoRa wireless communication is developed in~\cite{melo_low-cost_2021}. The data is sent to the LoRa Gateway by the data logger and forwarded to a MQTT Broker. The data is stored in the Google Cloud Platform. However, all of these devices are custom-built and cannot be considered widely available, which hinders widespread adoption.\\
\subsection{Implementation Challenges}
When integrating IoT devices into an electrical grid to improve the monitoring and control capabilities, a major challenge is network security~\cite{kimani_cyber_2019}. The potential number of devices in the grid creates a large attack surface. In addition, critical infrastructure is dependent on an uninterrupted supply of electricity, and attacks on the grid infrastructure could result in huge financial and economic losses~\cite{kimani_cyber_2019}. Energy infrastructure is therefore a popular target for cyber attacks. Attacks targeting IoT devices in the electricity grid include Denial-of-Service, Man-in-the-Middle, and Phishing attacks, whereas the latter two types of attacks being easier to execute when communications are not encrypted. Therefore, we encrypt the network communication of the smart plugs in our proposed approach.\\
An exemplary communication and data management platform is outlined in~\cite{bao_microservice_2016}, which collects measurement data from various devices through a variety of interfaces. It also adapts the protocols and stores the abstracted data in a database. The abstraction layer enables the unification of data collected by measurement devices from different vendors and their evaluation in new domains. In addition, by integrating multiple interfaces, it is possible to support multiple versions of software without having to worry about updating interfaces and losing support for older versions of software. To promote the reuse of functionality in~\cite{bao_microservice_2016}, auxiliary services are divided into components that are as small as possible. This approach follows the microservice philosophy. Moreover, a microservice-based architecture has the advantages of scalability, autonomy, and rapid deployment of new features~\cite{jamshidi_microservices_2018}. The lightweight technologies that a microservice relies on accelerate the development and deployment process. Deployment on servers using containerization leads to great autonomy of individual services and allows for dynamic allocation of resources~\cite{ibm_cloud_education_containerization_2021}. These advantages also lead to the increasing popularity of containerization of applications~\cite{ibm_cloud_education_containerization_2021}. To make our ICT structure for data collection and aggregation extensible, we decide to separate the different services and implement them as microservices. This allows us to support other vendors and communication channels without compromising compatibility with the technologies we already support. For example, the implementation of a REST API for data aggregation or the support of other database systems can be realized by implementing independent adapters.
\section{Method}
Smart plugs are plugs that connect to a Zigbee hub, a LoRaWAN hub, or a WiFi access point. They consist of an outlet that can be turned on and off by smartphone apps or a smart home hub. In addition to a relay to control the outlet and a microcontroller, some smart plugs also contain hardware to measure the power consumption of the connected device and the line voltage of the socket outlet they are plugged into. The measurement data is typically sent to a server of the device manufacturer, allowing customers to monitor the values measured by the smart device via a web service.
However, with suitable firmware, some smart devices can connect to IoT gateways other than the manufacturer's server. These gateways can forward the measured data, packaged into standardized messages, to a message broker, thus enabling remote monitoring and logging of the voltage levels and power consumption of connected devices. The data can then be stored as a time series and be used as input data for grid simulations, or be used in real-time to detect congestion, faulty hardware, or to control DSM hardware and distributed generation.

\begin{figure}[h]
\vspace{-8mm}
    \centering
\includegraphics[page=4]{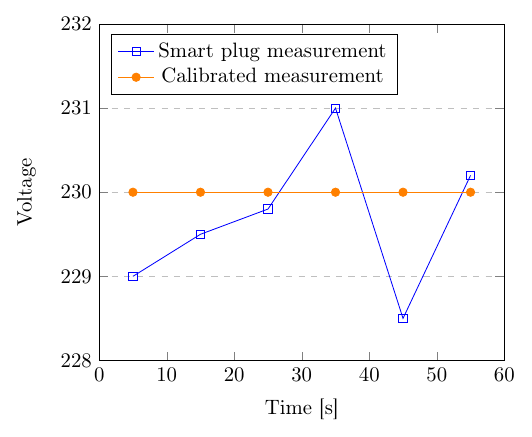}
    \caption{Smart plug measurements compared to voltage levels measured by a calibrated device}
\label{fig:accuracy}
\vspace{-8mm}
\end{figure}
Because these smart home devices are not intended to monitor the grid, the manufacturers of these devices do not provide data on the accuracy of the energy measurements. Furthermore, the accuracy and frequency of the measurements can be modified by modifying the firmware of the smart plugs. We analyze the measurements of the smart plugs with the unmodified firmware and our modified version. The difference between the values measured by a calibrated measurement device and the values measured by the smart plugs is reported as the accuracy of the smart plug. We use the measurements of a Janitza~UMG~604EP-PRO power analyzer as a calibrated reference. The measurement error shall not include a constant systematic bias, as we are treating this constant measurement error separately. The measurement error may be due to a change in temperature, interpolation between two measurement points, or insufficient resolution at some point in the measurement process. An example is shown in Figure~\ref{fig:accuracy}, where the difference between the calibrated voltage readings and the smart plug measurement is the measurement error.\\
In addition, smart home devices may be calibrated differently. Due to manufacturing tolerances and environmental differences between the smart home devices, the measured voltage and current levels can vary between devices from the same manufacturer and production batch. We calculate a constant offset bias for all smart plugs separately and remove this offset for each individual smart plug in a pre-processing step of the measured values. This calibration step must also be completed prior to deployment in a live environment.

\subsection{Communication Interface}
There are several types of smart home devices available. The main difference is the communication interface available, which can be based on WiFi, LoRaWAN or Zigbee. All three interfaces have different strengths and weaknesses as can be seen in the Table~\ref{tab:networking_technology}.

\renewcommand{\arraystretch}{2.5}
\setlength{\tabcolsep}{20pt}

\begin{table}[h]
\centering
    \begin{tabular}{c|c|c|c}
        Protocol & Hub needed & Data Rate & Range \\
        \hline
        WiFi & no & high & low \\
        LoRaWAN & yes & low & high \\
        Zigbee & yes & medium & medium \\
        
    \end{tabular}
    \vspace{4mm}
    \caption{Comparison of smart home communication technologies \cite{song_internet_2017}}
    \label{tab:networking_technology}
\end{table}

For this work, we use WiFi smart plugs. With a customized version of the Tasmota open source firmware~\cite{theo_arends_et_al_tasmota_2022}, it is possible to collect measurements every second and send them directly to a MQTT broker. With further modifications, a slightly higher measuring frequency could be realized, but this caused problems during practical tests. Compared to the LoRaWAN and Zigbee smart devices, no hub or gateway device is required other than the WiFi access point. In the test environment, access points are already in place, so no additional hardware is required, making the deployment of WiFi smart plugs the most practical option. In addition, the higher bandwidth allows for more frequent and comprehensive measurement data. With the smart plug used for testing in this paper, the measurements are available in the simulation in less than one second.\\
In general, the data sent over WiFi to an access point is not necessarily encrypted. However, the smart plugs evaluated in this paper contain an ESP8266 microcontroller that supports the WPA2 encryption standard. This enables the encryption of the communication between the smart plug and the access point, which protects the transmission of measurements.\\
The TLS encryption standard is supported by the Mosquitto MQTT broker we use, and the Tasmota firmware for the smart plug also includes basic support for this standard. To enable the ESP8266 to send TLS encrypted packets, a custom version of the Tasmota open source firmware must be compiled that includes the very lightweight BearSSL library. Since the smart plugs are configured with the SSL fingerprint of the MQTT broker and a preshared key, a spoofing attack in which the attacker impersonates the smart plug and sends malicious or false data is not trivially possible.\\
\begin{figure*}[h]
    \centering
\includegraphics[page=42,scale=1.2]{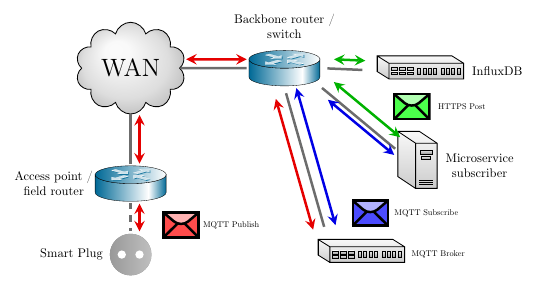}
\caption{Network infrastructure between the smart plug and the InfluxDB server}
    \label{fig:network}
\end{figure*}
A microservice application is used to subscribe to the MQTT broker. The application creates the adapter between the MQTT messages and the InfluxDB server. Incoming measurement data is mapped to specified fields. This architecture also allows for multiple different measurement devices to write to the same database server, and, in this case, to compare voltage measurements recorded by different devices. In addition, metadata can be added to the measurements so that the voltage data is associated with a power phase, a geographic location, and the manufacturer of the device.\\
This infrastructure also enables fast integration of new sensors by developing new microservices to map the measurement data messages. Should new smart plugs be introduced that do not support the MQTT protocol, new microservices can be added to inject data into the time-series database without losing support for the existing devices. Furthermore, the addition of other database servers for specific measurement data only requires the development of another microservice and does not require changes to the existing structures. An overview of the resulting networking infrastructure is shown in Figure~\ref{fig:network}.

\subsection{Measurement Hardware}
Besides the communication interface, another difference between the smart plugs is the measurement hardware. Popular energy measurement integrated circuits (ICs) for smart plugs are the Shanghai Belling BL0937 and the Shanghai Belling BL0940 \cite{tasmota_tasmota_2022}. The smart plugs used for testing in this paper contain the BL0937~IC.\\
The smart plugs used to perform the tests are numerous Nous A1T, Gosund SP1 and Shelly Plug S. However, the same measurement hardware is also included in many other smart plugs, so the measurement accuracy of the plugs is identical. All tested smart plugs share the same measurement characteristics.\\
The measurements taken by the smart plugs are compared to the values measured by a Janitza~UMG~604EP-PRO power analyzer. This power analyzer implements a measurement process according to IEC~61000-4-30 and is connected to an Influx database via TCP/IP.
%Since the manufacturers of the smart plugs do not specify the accuracy of the measurements, we have to collect and evaluate the measurement data ourselves. To do this, smart plugs are installed in a real world test environment. The smart plugs collect voltage measurements at the point of installation every ten seconds with the unmodified Tasmota firmware and every second with the modified variant. This data is then compared to voltage levels measured by a power analyzer device that is installed on the same phase. The difference between the measured values of the smart plug and the power analyzer is used to estimate the standard deviation of the smart plug measurements. Furthermore, the difference can be used to calculate a measurement offset due to the manufacturing tolerances of the smart plug. To do this, we assume that this offset is constant and calculate the average difference between the measured values of the smart plug and the power analyzer. This average is our offset value.

\section{Evaluation}\label{evaluation}
\begin{figure}[h]
    \centering
\includegraphics[page=47]{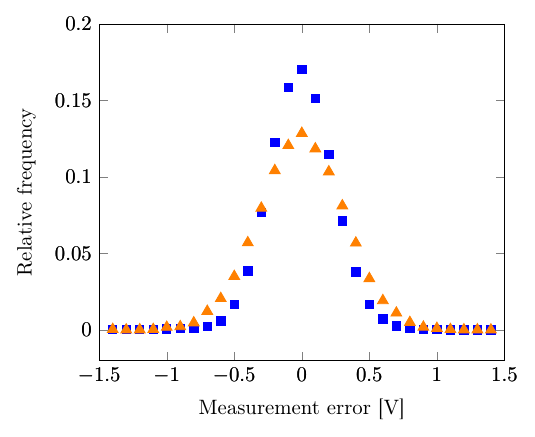}
    \caption{Relative frequency histogram of the measurement error of smart plugs with the modified firmware (blue) and the unmodified firmware (orange).}
\label{fig:error_hist}
    \vspace{-4mm}
\end{figure}

To evaluate the accuracy of the smart plug voltage measurement, two smart plugs are installed in the real-world test environment. In our test setup, the power analyzer is configured to send one measurement value per second. Since the smart plugs contain the same measurement IC, the difference in the measured values is only due to the difference between the modified and the unmodified firmware versions.\\
First, we analyze the measurements of the smart plug with the unmodified Tasmota open source firmware~\cite{tasmota_tasmota_2022}. With this firmware, the smart plugs output voltage measurements with one decimal place. Therefore one could assume that the error of a measurement is at most 0.1~V. However, due to rounding errors in the unmodified Tasmota firmware, the measurement error is higher. The smart plug only takes voltage measurements in steps of at least 0.2~V, sometimes even only 0.3~V.\\
In Figure~\ref{fig:error_hist}, the measurement error of the smart plug with the unmodified firmware version is plotted in orange and the measurement error of the smart plug with the modified firmware is plotted in blue. Since the unmodified firmware version outputs one measurement value every ten seconds and the modified firmware version outputs one value per second, there are exactly ten times as many measurement values of the modified firmware version in the same time span. The Y-axis values are relative to the total number of measurements.\\

The measurements of the smart plug with the unmodified Tasmota firmware are more spread out compared to the measurements of the measurements with the modified firmware, indicating that the standard deviation of the blue measurements is lower than the standard deviation of the orange measurements. This is indeed the case, as the standard deviation of the smart plug with the unmodified firmware is 0.33~V and the standard deviation of the smart plug with the modified firmware is 0.27~V.\\

Neither the Anderson-Darling test \cite{m_a_stephens_edf_1974} for normality nor the Shapiro-Wilk test \cite{shapiro_analysis_1965} allow us to reject the null hypothesis that the data are normally distributed. The Anderson-Darling test returns a statistic of 0.44 and a critical value to reject the null hypothesis of 0.57, even at a significance level of 15~\%. The Shapiro-Wilk test gives a p-value of 0.54. Therefore, we conclude that the measurement error is likely to follow a normal distribution or some other very similar distribution.\\
\section{Case Study: Monitoring the Distribution Grid with Smart Plugs}
The use of smart plugs as distributed measurement devices in the distribution grid can enable real-time monitoring of voltage levels, allowing utilities to detect and address issues promptly. However, the measurement error could be an issue for accurate distribution grid monitoring. To further explore the impact of the measurement error and the correlation between the number of measurement devices and the accuracy of the grid monitoring, this section presents a case study of an exemplary distribution grid monitoring.\\
In the case study, we simulate the power flow in a IEEE 37 bus system. The smart plugs installed in the grid are simulated as well with a measurement error according to our findings in the previous section. An integration of real world measurements into a power flow simulation is not shown in this case study.

\subsection{Problem formulation}
Distribution grid monitoring and state estimation are becoming increasingly important for DSOs due to the rise in flexible consumption, distributed generation and the increase of demanding loads such as heat pumps and electric vehicle chargers. However, accurately monitoring the distribution grid and determining the impact of new loads and distributed electricity generation requires a large number of measurement devices. This case study shows how a limited number of smart plugs can provide valuable insight into the voltage levels at different nodes within the distribution grid area. It also outlines the relationship between the number of smart plugs in the grid area and the accuracy of the monitoring.

\subsection{Method}
\begin{figure}[h]
    \centering
\includegraphics[page=19]{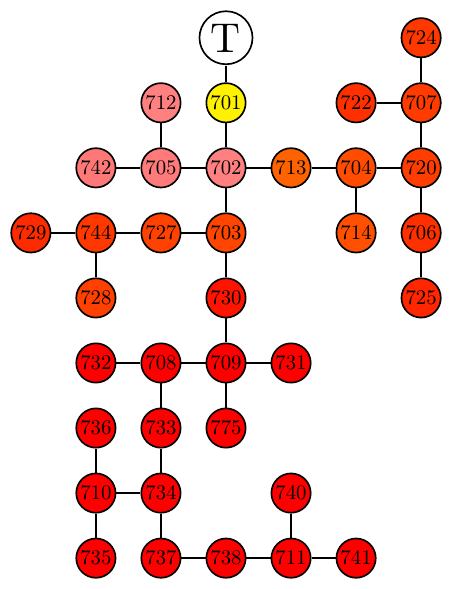}
\caption{IEEE bus system 37. Nodes are colored based on the voltage monitoring error. Monitored voltages at white and yellow nodes are similar to the simulated voltages and monitored voltages at the red nodes differ more from the simulated voltages. In this instance, only one smart plug is installed at node 736 and the average monitoring error is about 2.87~V}
    \label{fig:bus_system_37}
    
    \vspace{-4mm}
\end{figure}

To evaluate the benefit of smart plug measurements for grid state monitoring, an IEEE 37 bus system is simulated. We implement the grid simulation using Pandapower, an open-source tool written in Python for modeling and analyzing of power grids~\cite{thurner_pandapoweropen-source_2018}. The smart plugs providing the measurements are also simulated. Figure~\ref{fig:bus_system_37} shows an overview of this standardized distribution grid bus system.

The transformer \textbf{T} is connected to the 20~kV grid on the primary side and to the 400~V distribution grid on the secondary side. The nodes in the graph represent the houses in the distribution grid. In this power flow simulation, all the houses are placed 40~m away from each other, and NAYY~4x150~SE lines are used to connect them. These are the most common lines used in Germany~\cite{kochanneck_systemdienstleistungserbringung_2018} and 40~m is a common distance between neighbors in a rural German distribution grid~\cite{georg_kerber_statistische_2008}. We assume that the root mean square (RMS) voltage at the transformer is constant for this simulation model.\\
To evaluate the impact of the measurement error of the distributed smart plugs on the monitoring of the grid area, we implement several scenarios with different numbers of smart plugs installed in the grid area. We also compare the monitoring results that are based on the smart plugs with the unmodified firmware with the results based on the smart plugs with the modified firmware version.\\
All scenarios are based on the IEEE bus system 37. For each node in the original bus system, we add and connect another node representing the house and the smart plug inside the house. We also add random resistive loads between 0~kW and 6.5~kW with a power factor of 1.0 to the house nodes, representing household appliances and electric vehicle chargers. The DSO sees the total load at the feeder, but does not know where the individual loads are located. The simulated voltage levels are the ground truth against which we will later compare our monitoring results.\\
The DSO monitors the total load at the feeder level and the voltage level at the outlets where the smart plugs are located. In the case study, we use the voltage levels generated by the simulation as the ground truth. We offset them with random values sampled from a normal distribution with the standard deviation calculated in section~\ref{evaluation} to model the measurement accuracy of the smart plugs. This results in voltage levels that could be measured by a DSO in a real world experiment and we call them artificial voltage measurements.\\
We now attempt to estimate the true voltage levels at all nodes in the bus system 37 from the perspective of the DSO based on the artificial voltage measurements and the feeder load. To do this, we calculate the average load by dividing the feeder load by the number of houses in the grid. We then place this average load at each node in the grid as a starting point. This results in a rough estimate of the voltage levels at all the nodes.\\
Next, we compare the voltage levels at each house to the artificial voltage measurements recorded by our artificial smart plugs and approximate the loads at these measurement points. If the voltage at a house node is higher than the artificial voltage measurements, the assigned load is increased. If the voltage is lower, the load is reduced. The result is a distribution grid with loads placed at all nodes so that the voltage levels measured by the smart plugs match the true voltage levels. However, due to the inaccuracy we intentionally introduced into the simulated voltage levels, these loads and voltage levels differ from the ground truth.
\subsection{Evaluation}
\begin{figure}[h]
\vspace{-4mm}
    \centering
\includegraphics[page=40]{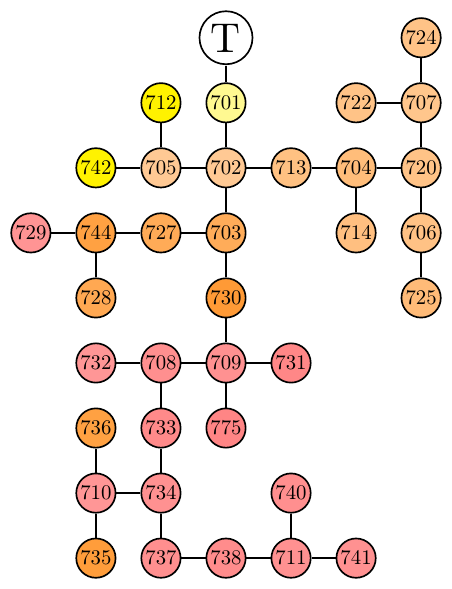}
\caption{IEEE bus system 37. Nodes are colored based on the voltage monitoring error. Monitored voltages at white and yellow nodes are similar to the simulated voltages and monitored voltages at the red nodes differ more from the simulated voltages. In this example, eight smart plugs are installed at nodes 736, 706, 709, 711, 742, 722, 725 and 738. The average monitoring error is about 1.04~V and the coloring is consistent with Figure~\ref{fig:bus_system_37}}
    \label{fig:bus_system_37_8sp}
    \vspace{-4mm}
\end{figure}

In order to evaluate the effect of distribution network metering devices on the quality of monitoring, we compare the artificial voltage measurements generated as described above with the true voltage levels. The nodes in Figure~\ref{fig:bus_system_37} are colored based on the difference between the true voltage levels and these artificial voltage measurements. Red nodes represent a greater difference between the true voltage levels and the artificial voltage measurements and the lighter the nodes are colored, the smaller the monitoring error is. The voltage error at the feeder is the smallest . This is due to the short length of the line and the small voltage drop across the line between the feeder and the first house. In the grid shown in Figure~\ref{fig:bus_system_37} only one smart plug is used for monitoring.
\newpage
In Figure~\ref{fig:bus_system_37_8sp}, eight smart plugs are placed in the grid area. It can be clearly seen that the monitoring error is lower at each node in the grid.

\begin{figure}[h]
    \centering
\includegraphics[page=44]{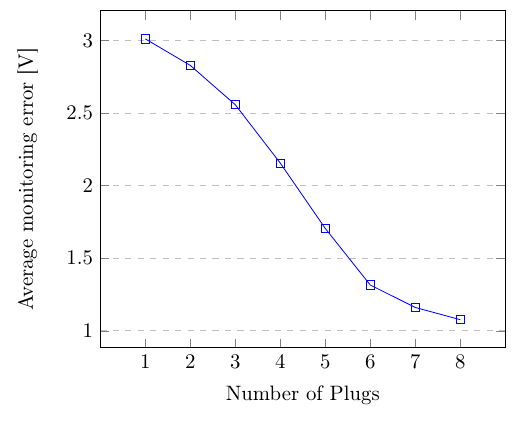}
\caption{Correlation between the number of smart plugs in the grid area and the average monitoring error.}
    \label{fig:num_plugs_error}
    \vspace{-4mm}
\end{figure}

The correlation between the number of measurement devices in the grid and the average voltage error in a 230~V grid can be seen in Figure~\ref{fig:num_plugs_error}.
%
%\section{Local Household Load Monitoring with Smart Plugs}
%\subsection{Problem formulation}
%Privacy concerns
%\subsection{Methodology}
%Machine Learning
%\subsection{Evaluation}
\section{Discussion}
The voltage standard deviation observed in the smart plug measurements with the modified firmware of 0.27~V is well within the range of what is considered acceptable in other publications (0.6~\% in~\cite{samarakoon_use_2011} and 0.3~\% to 0.9~\% in~\cite{jia_state_2013}). Installing multiple smart plugs in the same distribution grid area further improves the accuracy of the measurements, and monitoring multiple phases in three-phase distribution networks would allow asymmetric loads to be detected. This should be evaluated in the future. The smart plugs with the modified firmware version allow for up to one voltage measurement per second, which are available almost immediately for a grid state analysis. In contrast, smart meters often take only one measurement every fifteen minutes and transmit the data at intervals of up to six hours~\cite{sanchez_observability_2017}. In addition, the deployment of the smart plugs in a real-world test environment can be completed in minutes by configuring the smart plug and connecting it to a nearby WiFi network, and no electrician is required for installation.\\
Smart meters are typically installed near to the point of common coupling. Smart plugs, on the other hand, measure the voltage at the outlet to which they are connected to. This means that the voltage drop within the resident's home is included in the smart plug's measurements. This voltage drop depends on the loads within the home's electrical system and is therefore not constant. In order to reduce the voltage drop on the local line, it is necessary to install the smart plug as close as possible to the point of common coupling. In addition, the smart plugs monitor only one phase. However, the load in the distribution grid is predominantly symmetrical~\cite{schwab_elektroenergiesysteme_2017}, which means that the voltage drop is also symmetrical.\\
The approach of using widely available smart plugs to monitor the distribution grid is mostly limited by privacy concerns and the accuracy of the measurements, especially when compared to the measurements from calibrated smart metering systems or power analyzers. Installing the smart plugs away from the common coupling point further reduces the validity of the measurements, and the constant offset of each device must be determined and compensated for. However, the frequency of the measurements could permit some compensation for these shortcomings, e.g. by means of filters. Another issue for practical implementation in the field is the availability of a WiFi connection to transmit the measurements. Perhaps allowing customers to use the switching function of the smart plug would be an incentive to allow the use of their private WiFi access.\\
In general, the installation of a custom firmware to monitor the distribution grid voids the warranty of the smart plugs. The manufacturers of the smart plugs would need to provide a software interface to collect measurement data or connect to a custom server in order to use the smart plugs without flashing a custom firmware. Without the manufacturers support for such a feature, the DSO would need to flash the custom firmware before distributing the smart plugs to the customers.
\section{Conclusion}
In this paper, we determine the accuracy of smart plug measurements by comparing the calculated values with voltage readings from a Janitza~UMG~604EP-PRO power analyzer. We use commercially available devices in the present work that are able to connect directly to a WiFi network and transmit the measurement data to a server, eliminating the need for a relay. The voltage measurements of the tested smart plugs with the modified Tasmota firmware have a standard deviation of 0.27~V, which is lower than the standard deviation of the measurements taken by smart plugs with the unmodified Tasmota firmware. The modified firmware is published as open-source. We also describe the network structure and the integration of smart plug measurement data into an existing time-series database. In a case study, a practical use-case for a distribution grid monitoring is outlined and evaluated. It is shown how the average monitoring error in a distribution grid area decreases the more measurement devices are installed. The installation of the commercially available smart plugs does not require an electrician, the hardware is inexpensive, and the individual configuration of the devices is simple. In this light, simple smart plugs can help with distribution grid monitoring and provide valuable information to the DSO.\\
Future research should evaluate other use cases of smart plugs for a DSO, such as DSM and non-intrusive load monitoring. Furthermore, other types of power measurement ICs in other smart home devices should be evaluated and compared. To complete the monitoring and account for asymmetric loads, it would be relevant to consider two-phase or three-phase distribution systems. And in a real-world implementation of a grid monitoring using smart home devices, the placement of the devices in the grid must be optimized to collect the most useful data.
\newpage
\subsubsection*{\ackname}
This research has been funded by the German Federal Ministry for Economic Affairs and Climate Action (TrafoKommune project, funding reference: 03EN3008F)
\bibliographystyle{splncs04}
%\bibliography{references}

%
\end{document}